\documentclass[reprint]{revtex4-1}

\usepackage{graphicx, amsmath, color, upgreek}

\begin{document}


\title{Measurement of a Magnonic Crystal at Millikelvin Temperatures}
\author{S. Kosen} \email{sandoko.kosen@physics.ox.ac.uk}
\author{R.\,G.\,E. Morris}
\author{A.\,F. van Loo}
\author{A.\,D. Karenowska}
\affiliation{Clarendon Laboratory, Department of Physics, University of Oxford, Oxford OX1 3PU, United Kingdom}

\date{\today}

\begin{abstract}
Hybrid systems combining magnons and superconducting quantum circuits have attracted increasing interest in recent years. Magnonic crystals (MCs) are one of the building blocks of room-temperature magnonics and can be used to create devices with an engineered band structure. These devices, exhibiting tunable frequency selectivity and the ability to store travelling excitations in the microwave regime, may form the basis of a set of tools to be used in the context of quantum information processing. In order to ascertain the feasibility of such plans, MCs must be demonstrated to work at the low temperatures required for microwave-frequency quantum experiments. We report the measurements of the transmission of microwave signals through an MC at $20$\,mK and observe a magnonic bandgap in both continuous-wave and pulsed excitation experiments. The spin-wave damping at low temperatures in our yttrium iron garnet MC is higher than expected, indicating that further work is necessary before the full potential of quantum experiments using magnonic crystals can be realised.
\end{abstract}


\maketitle

Superconducting quantum circuits have become an increasingly mature experimental technology in recent years \cite{Devoret2004, Gambetta2017}. As a result, there has been a surge of interest within the circuit quantum electrodynamics (circuit QED) community in combining such circuits with other physical systems such as spin ensembles \cite{Kurizki2015,You2011}, acoustic waves \cite{Gustafsson2014}, and magnonic structures \cite{Tabuchi2016}.

The goal of quantum magnonics is to investigate the physics of magnons at the quantum level and to create novel microwave devices useful for quantum information processing. Dipolar magnons (spin waves) \cite{Kruglyak2010} have $\upmu$m-wavelengths and are readily excited over a range of microwave frequencies which overlap with those of superconducting quantum circuits. Recent work includes the measurement of surface spin waves in a $\upmu$m-thick yttrium iron garnet (YIG) waveguide at millikelvin temperatures \cite{vanloo2016}, the demonstration of strong coupling between bulk YIG samples and resonators \cite{Huebl2013, Tabuchi2014, Goryachev2014, Zhang2015, Bourhill2016, Kostylev2016, Morris2017} and the excitation of a single magnon in a YIG sphere using a superconducting qubit \cite{Tabuchi2015, Lachance-Quirion2017}.

Magnonic crystals (MCs) \cite{Krawczyk2014, Chumak2017}, the magnetic analogue of photonic crystals, are magnetic waveguides with artificially engineered magnonic bandgaps. MCs are created by imposing periodic changes in a waveguide's magnetic properties or environment. Various implementations have been demonstrated, including several static varieties and a dynamic variant with a bandgap that can be switched on and off.

At room temperature, MCs have been used to create a range of devices including oscillators and filters \cite{Karenowska2010}, logic gates \cite{Nikitin2015}, and magnon transistors \cite{Chumak2014}. Several of the properties of magnonic crystals --- notably their strong and tunable frequency selectivity, storage capability, and ability to alter the propagation direction of signals --- have potential utility in the manipulation of single magnon excitations in experimental solid-state quantum devices \cite{Chumak2010, Chumak2012}. Until now, however, it remained to be established that MCs can be used at the millikelvin temperatures required for such devices. In this work, we present measurements of a magnonic crystal at millikelvin temperatures, a step towards the incorporation of MCs into quantum devices.

\begin{figure}
\includegraphics{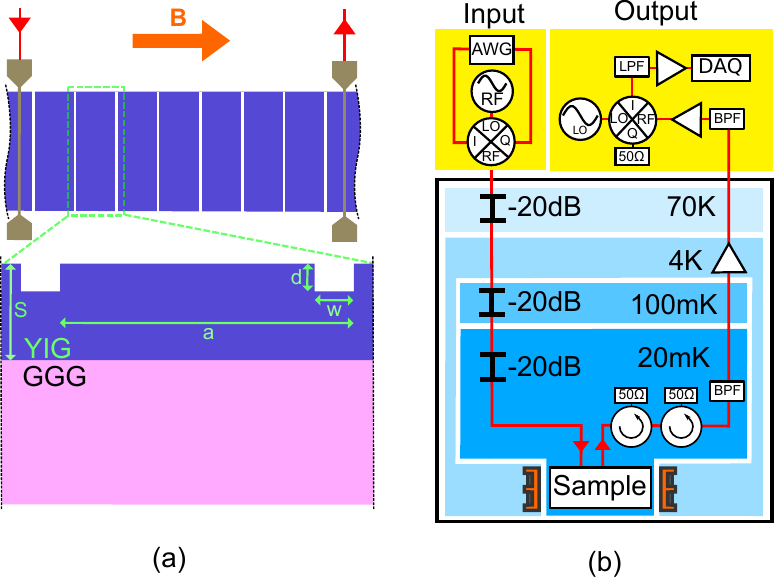}
\caption{(a) Simplified illustration of the magnonic crystal used in our experiments (not to scale). Here, $S=5.19\,\upmu$m, $a=300\,\upmu$m, $w=40\,\upmu$m, and $d=0.5\,\upmu$m. (b) Schematic of the measurement setup. \label{MC_fridge}}
 \end{figure}

The basis for the magnonic crystal used in our experiments is a structured YIG waveguide (thickness $S=5.19\,\upmu$m, room-temperature saturation magnetization $M_s=138.6$kA/m) epitaxially grown on a gadolinium gallium garnet (GGG) substrate. YIG, a ferrimagnetic electrical insulator, has extremely low spin-wave damping at room temperature and is therefore much used in room-temperature magnonic device development \cite{Serga2010}. The MC is formed from a series of eight equally-spaced grooves, each of width $w=40\,\upmu$m and depth $d=0.5\,\upmu$m, chemically etched into the magnetic film. The distance between the grooves is $a=300\,\upmu$m (see fig.\,\ref{MC_fridge}(a)). Spin waves are excited and detected by niobium microstrip antennae fabricated $2.66\,$mm apart on a sapphire crystal substrate in direct contact with the MC. In order to assure compatibility with the thin-film superconducting measurement structures used in circuit QED, it is desirable to apply the required bias magnetic field in-plane. We chose to carry out our experiments in the backward volume geometry (BVMSW) \cite{Hurben1995} (bias magnetic field parallel to the spin-wave propagation direction ($\vec{k}\parallel \vec{B}$), which is along the longitudinal axis of the waveguide). At room temperature, crystals measured in the backward volume configuration have been shown to display bandgaps with a higher rejection ratio than magnetostatic surface spin waves (MSSW) ($\vec{k}\perp\vec{B}$, in-plane field) \cite{Chumak2009}.

A dilution refrigerator is used to cool the MC assembly, housed in a copper sample box, down to $20\,$mK. Figure \ref{MC_fridge}(b) shows a schematic of the measurement setup.  A superconducting coil provides the static magnetic field $B$ necessary to bias the magnetic sample. A microwave source, IQ-mixer and arbitrary waveform generator (AWG) are used to create microwave input pulses. Three $20$\,dB attenuators in the input line ensure that the electrical noise temperature of the input signals is comparable to the thermodynamic temperature of the sample. Output signals pass through two $50\,\Omega$-terminated circulators before being amplified at the $4\,$K stage. Outside the fridge, the output signals are down-converted to 500\,MHz. A fast data acquisition (DAQ) card digitises the down-converted signal at a sampling frequency of 2.5\,GHz. Signals are typically averaged 20,000 times on the DAQ card before being digitally demodulated.

\begin{figure}
\includegraphics{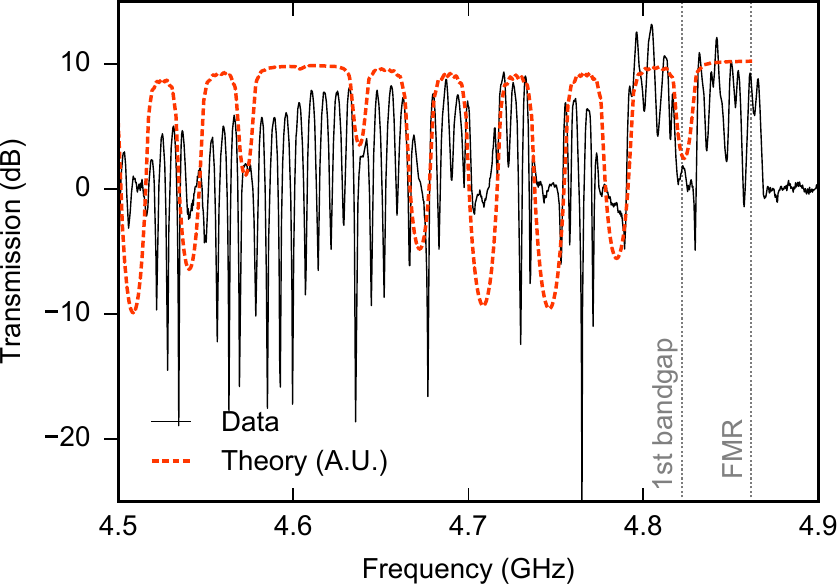}
\caption{Transmission of continuous-wave BVMSW signals through the magnonic crystal measured at room temperature using a network analyser (solid line). The bias magnetic fixed is fixed at $B=107$\,mT. Data is relative to that measured at zero-field, i.e. when no spin waves are excited and only directly-coupled electromagnetic signals contribute. The theoretical transmission (dashed line) is calculated using the transfer matrix method with $M_s=138.6$kA/m, $\eta=8$, $\zeta=0$, $\Delta H = 0.5$\,Oe.}\label{theory_S21_MC10}
\end{figure}

The MC is first characterised at room temperature using a network analyser. Figure \ref{theory_S21_MC10} shows the transmission measured at room temperature as a function of microwave input signal frequency with $B=107$\,mT. The displayed data is relative to that measured at zero field, i.e. when no spin waves are excited within the waveguide and only directly-coupled electromagnetic signals propagate between the input and output antennae through the vacuum of the sample box. In this figure, the highest frequency at which the BVMSW are observed corresponds to the spins precessing uniformly throughout the material (FMR, $k=0$). Propagating modes ($k\neq 0$) have lower frequencies. The low-frequency (high-$k$) limit of the measurable band is determined by the geometry of the microwave antennae. High-$k$ excitations couple less well than low-$k$ ones to these structures, the coupling becoming negligible once the wavelength is smaller than the antenna width.

Below the FMR frequency, the data displays oscillations caused by the interference between the spin-wave signal and the directly-coupled signal; due to the different dispersion relation of the magnonic and photonic waves, these signals accumulate different phases while travelling to the output antenna, resulting in interference fringes. The magnonic bandgaps of the crystal appear as gaps in this pattern: in the bandgaps, the transmitted spin-wave signal is suppressed while the directly-coupled signal is unaffected, resulting in regions without oscillations.

Calculations were made using the transfer matrix method following the treatment of Chumak \cite{Chumak2009}. In this model, spin waves accumulate phase and experience damping while propagating in between neighbouring edges of the grooves defining the lattice of the magnonic crystal. At the interfaces between etched and unetched regions, spin waves undergo partial reflection and transmission. For completeness, it should be noted that the coupling of the antennae to the waveguide has some dependence on $k$ which is not included in the model: the effect of this on the key qualitative features being fitted (namely the position and width of the bandgaps) is negligible.

\begin{figure}
\includegraphics{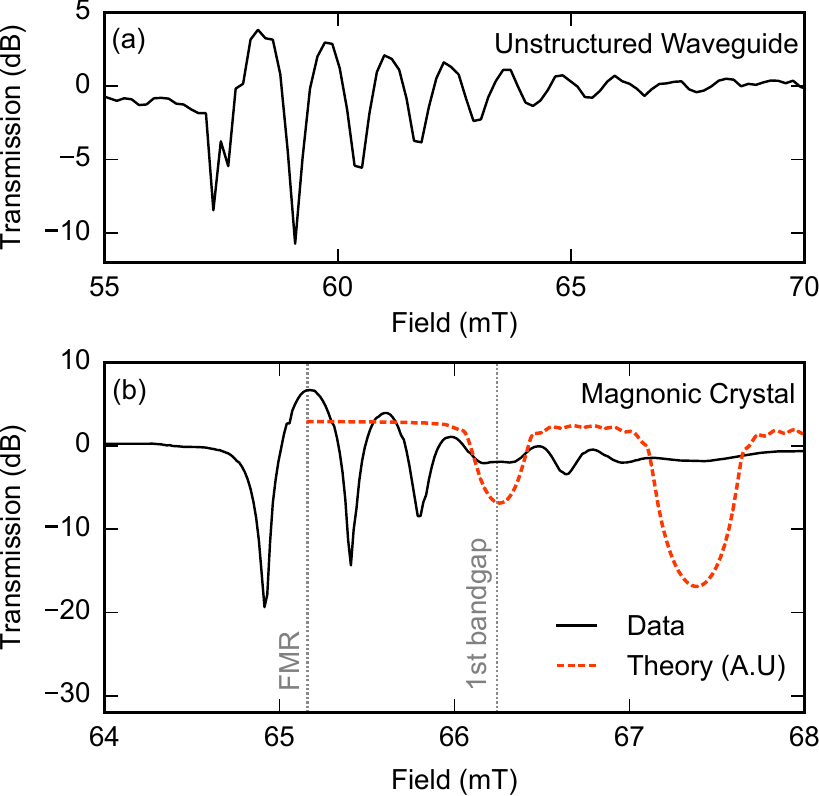}
\caption{Transmission of continuous-wave BVMSW signals measured at $20\,$mK. Measurements are performed by sweeping the magnetic bias field ($B$) while applying a $4\,$GHz microwave input tone through (a) an unpatterned magnonic waveguide ($11\,\upmu$m thickness, $1\,$mm antenna separation), and (b) a magnonic crystal waveguide ($5.19\,\upmu$m thickness, $2.66\,$mm antenna separation). An offset has been applied to the data to shift the baseline to 0 dB. The theoretical curve is calculated using the transfer matrix method with $M_s=197$\,kA/m, $\eta=8$, $\zeta=0$, $\Delta H = 0.5$\,Oe.} \label{normal_MC10_10mK_theory}
\end{figure}

Apart from the FMR linewidth ($\Delta H$), two phenomenological parameters appear in the model: $\zeta$ which accounts for the increased damping due to two-magnon scattering within the grooves, and $\eta$ which is used to match the predicted and observed width and depth of the bandgaps. For simplicity, in our calculations $\zeta$ is set to zero and $\eta$ is adjusted to fit the measured widths of the gaps. The theoretical prediction of the transmission characteristics across the magnonic crystal with $M_s=138.6\,$kA/m (dotted line in fig.\,\ref{theory_S21_MC10}) is consistent with the observed positions and widths of the bandgaps. 

Figure \ref{normal_MC10_10mK_theory} compares the transmission characteristics of an unstructured magnonic waveguide  ($11\,\upmu$m film thickness, $1\,$mm inter-antennae spacing) and the same magnonic crystal at 20\,mK. An offset has been applied to the data to shift the baseline to $0$\,dB. In contrast to the room-temperature measurement in fig.\,\ref{theory_S21_MC10}, measurements at $20\,$mK are made as a function of the magnetic bias field ($B$) while keeping the input frequency constant. The system is excited using a constant frequency 4\,GHz microwave tone with a power of $-70\,$dBm at the input of the antenna. The lowest field at which the BVMSWs are observed corresponds to the FMR. Signals at higher fields are propagating modes ($k\neq 0$).

At $20\,$mK, the measurement of the unstructured waveguide (fig.\,\ref{normal_MC10_10mK_theory}(a)) shows oscillations across the spin-wave passband that decay in amplitude as $k$ increases (i.e. as $B$ increases). As in the data of fig.\,\ref{theory_S21_MC10}, the oscillations are due to the interference between the spin-wave and directly-coupled signals. As anticipated, without the etched grooves, no magnonic bandgap is present. In the MC measurement (fig.\,\ref{normal_MC10_10mK_theory}(b)), a single bandgap is observed. Its position at $\sim$\,66.2\,mT agrees with that predicted using the transfer matrix method with a saturation magnetisation of $M_s=197$\,kA/m \cite{vanloo2016, Maier-Flaig2017}. Note that in the transfer matrix model, neither the position nor the width of the bandgaps are significantly affected by changing the parameter $\Delta H$. Accordingly, when we calculate where the bandgaps are expected at low temperature, we use the room-temperature value of this parameter with the proviso that this modelling is not intended to reveal the additional low-temperature damping: it is solely a tool for identifying the position and width of the gaps.

The range of $k$-values over which spin-wave signals are measurable is found to be substantially narrower at low temperature than it was at room temperature ($\sim168$\,rad/cm at $20\,$mK versus $\sim1060$\,rad/cm at room temperature), indicating a higher damping constant. The effect of higher damping on the measured signal is more severe at higher $k$ (higher field in fig.\,\ref{normal_MC10_10mK_theory}, lower frequency in fig.\,\ref{theory_S21_MC10}) owing to the shape of the BVMSW dispersion curve: since the magnitude of the spin-wave group velocity decreases with increasing wavenumber, excitations with higher $k$ take longer to traverse the waveguide and are therefore more severely damped \cite{Chumak2009}. In our $20$\,mK experiments, the spin-wave signal at the $k$-value corresponding to the second bandgap is too weak to be detected.

\begin{figure}
\includegraphics{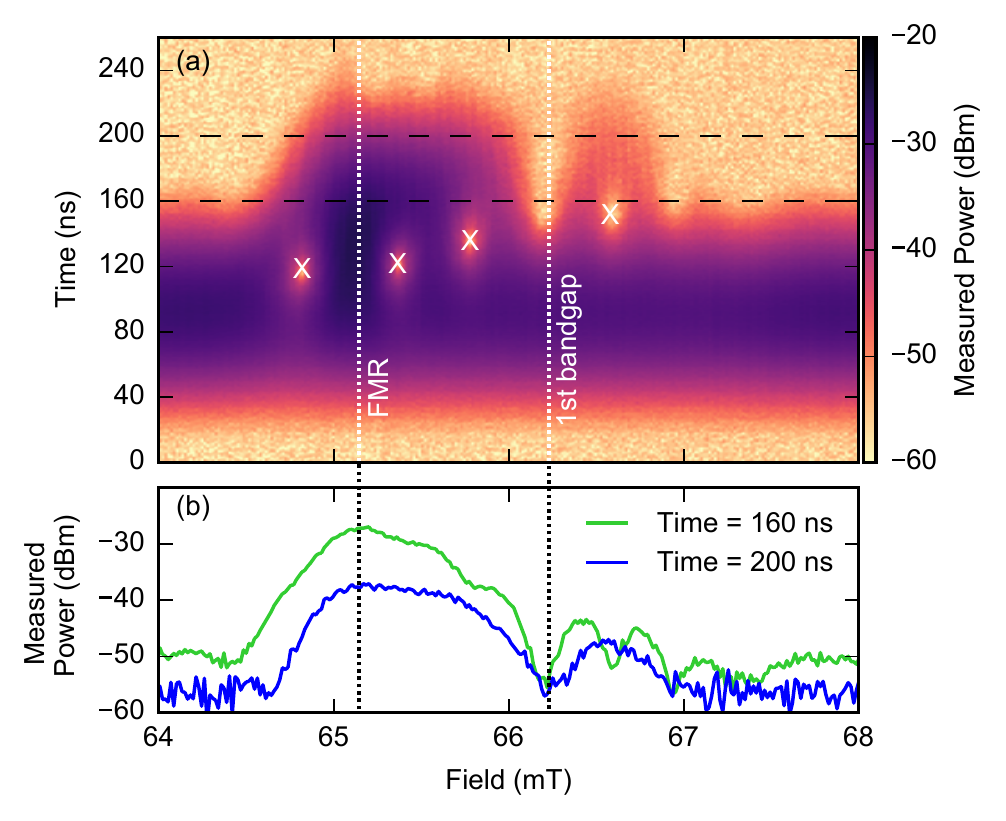}
\caption{Time-resolved measurements of pulsed BVMSW signals through the magnonic crystal at $20\,$mK. (a) Measurements are performed by sweeping the bias magnetic field $B$ while applying a $4$\,GHz microwave tone with a Gaussian envelope ($\sigma=30$\,ns). The dominant measured response in the horizontal band from $\sim40\,$ns to $\sim160\,$ns is the directly-coupled signal. The markers `X' indicate where the spin-wave and the directly-coupled signals interfere destructively. Beyond {160}\,ns, the directly-coupled signal disappears, leaving only the transmitted spin-wave signal. A bandgap is observed at $66.2\,$mT. (b) Linecuts at $t=160$\,ns and $t=200$\,ns as a function of magnetic field.}\label{combined_gaussianpulse_30ns}
\end{figure}

The presence of a magnonic bandgap can also be observed in time-resolved measurements at 20\,mK. Since the spin waves propagate slower than the directly-coupled signal (which travels at the speed of light to the output antenna), for sufficiently short excitation pulses, the two can be separated in time. Care has to be taken, however, not to make pulses so short as to have a frequency bandwidth that exceeds the width of the bandgap: under these conditions the gap cannot be observed since the signal always has a component lying outside of the bandgap which can propagate freely through the crystal. 

Figure \ref{combined_gaussianpulse_30ns} shows the time response of the magnonic crystal to a Gaussian pulse ($\sigma=30\,$ns) with a carrier frequency of $4$\,GHz. Such pulses are slightly too long to allow complete temporal separation between the directly-coupled signal and the spin-wave signal, but the bandgap starts to be obscured if they are made shorter. Initially, only the directly-coupled signal is measured ($\sim40$\,ns $< t <\,\sim 100$\,ns). When the spin waves start to arrive at the output antenna ($\sim100$\,ns $< t<\,\sim160$\,ns), they overlap in time with the directly-coupled signals, interfering destructively at `X' in fig.\,\ref{combined_gaussianpulse_30ns}. Beyond 160\,ns, the directly-coupled signal disappears, leaving only the transmitted spin-wave signal. Figure \ref{combined_gaussianpulse_30ns}(b) shows the linecuts from the same data at $t=160$\,ns and $t=200$\,ns. The first bandgap of the magnonic crystal is visible at $66.2$\,mT, consistent with the continuous-wave measurement in fig.\,\ref{normal_MC10_10mK_theory}(b).

A comparison between the room-temperature (fig.\,\ref{theory_S21_MC10}) and cold data (fig.\,\ref{normal_MC10_10mK_theory}(b)) indicates the presence of a significant increase in spin-wave damping at millikelvin temperatures. There are three possible sources of damping that warrant careful consideration: magnetic impurities in the YIG, enhanced damping due to the scattering processes caused by uneven etching of the grooves, and the GGG substrate upon which the MC is grown.

Previous measurements \cite{Spencer1959, Maier-Flaig2017, Haidar2015, Jermain2017} have shown that FMR linewidths in YIG initially increase as the material's temperature is decreased (below $100$\,K), reach a maximum value, and then begin to reduce again. This is generally attributed to the presence of paramagnetic rare-earth impurities in YIG with temperature-dependent relaxation times. While the lowest temperatures reached in these earlier works are around $5\,$K, they consistently report decreasing linewidths when the temperature is reduced below $10\,$K. Furthermore, the linewidths of YIG spheres measured in Refs.\,\cite{Tabuchi2014} and \cite{Zhang2015} at millikelvin temperatures are similar to the values observed at room temperature. From this, it seems likely that it is feasible to produce a pure YIG material with a linewidth at millikelvin temperatures comparable to the room-temperature value.

The surface roughness of a ferrite sample is known to influence the FMR linewidth because it increases two-magnon scattering, especially in a thin film sample \cite{Gurevich1996}. Spencer \cite{Spencer1959} has shown that better polished YIG spheres do exhibit lower linewidths across a range of temperature, from $300\,$K down to $5$\,K. Rough surfaces inside the grooves which define an etched MC are known to contribute to damping \cite{Chumak2009} but, as yet, there is no reason to think that this effect would be significantly enhanced at low temperatures.

The substrate upon which the YIG film is grown, gadolinium gallium garnet, is known to be paramagnetic below $70$\,K \cite{Danilov1989}. GGG is well-known to have a frustrated spin system with an ordered antiferromagnetic state below $400\,$mK at a relatively high field ($\sim 1\,$T) \cite{Schiffer1994}. At low field, the material undergoes a spin glass transition below $\sim200\,$mK \cite{Schiffer1995}. While its behaviour at the intermediate field ranges of our experiments is not well-documented, given these known magnetic properties and the relatively narrow linewidths measured in bulk YIG at low temperature (i.e. in the absence of GGG) it seems highly likely that, if not the only culprit, losses due to its low-temperature magnetic system coupling to the YIG are at least an important contributor to the increased damping we observe.

In conclusion, we have measured a bandgap in a magnonic crystal consisting of an etched YIG waveguide at $20$\,mK. Our results are consistent with calculations based on the transfer matrix method, both for continuous-wave and time-resolved measurements. Room-temperature and cold measurements of the same magnonic crystal indicate the presence of higher-than-expected spin-wave damping in the YIG at millikelvin temperatures. Future experiments investigating spin waves in YIG waveguides at millikelvin temperatures may provide more insight into the nature of this damping. This is essential if magnonic crystals are to be used for manipulation of magnons at the quantum level.

\begin{acknowledgments}
This work was supported by the Engineering and Physical Sciences Research Council grant EP/K032690/1. We acknowledge A.V.\,Chumak for the magnonic crystal sample and J.F.\,Gregg for the use of his room-temperature magnet. S.\,Kosen acknowledges the Indonesia Endowment Fund for Education.
\end{acknowledgments}

\bibliography{MCbib}

\end{document}